# Hybridization Effects in Ni-Mn based Shape Memory Alloys: XAFS Study

K. R. Priolkar[*], P. A. Bhobe, P. R. Sarode

Department of Physics, Goa University, Taleigao Plateau, Goa 403 206 India

krp@unigoa.ac.in

**Abstract:** Martensitic and magnetic properties of ferromagnetic shape memory alloys are known to depend up on structural modulations and associated changes in the Fermi surface. These modulations although periodic and spanning over multiple unit cells, involve movement of atoms typically of the order of 0.01Å. Therefore X-ray Absorption Fine Structure (XAFS) is an ideal tool to map both, local atomic movements and changes in density of states (DOS) due to changing hybridization as the system transforms from austenitic to martensitic phase. This paper presents a compilation of our XAFS studies on the Ni-Mn based shape memory alloys. A complete description of the changes in local structure around the constituent metal ions in the following alloy compositions: $Ni_{2+x}Mn_{1-x}Ga$, $Ni_2Mn_{1.4}Sn_{0.6}$ and $Ni_2Mn_{1.4}In_{0.6}$ in the austenitic and martensitic phases have been obtained. The results give the new experimental evidence for the crucial hybridization component that influences and leads to structural transition in these Ni-Mn based Heusler alloys.

**Introduction**

Martensitic transformations and its pairing with ferromagnetism has been a central subject for investigation in the recent years. Especially, some intermetallics show simultaneous occurrence of martensitic and magnetic transitions, suggesting the possibility of controlling the structural transformation by magnetic field and could be exploited for practical applications. Such multifunctional materials are classified under rapidly growing technological field of Ferromagnetic Shape Memory alloys (FSMA). Among the variety of FSMA, Ni-Mn-Ga alloys are a recently synthesized class of alloys that have been studied extensively and hence serve as a reference in the development of new systems [1, 2, 3]. The stoichiometric $Ni_2MnGa$ undergoes martensitic transition around 220 K from a $L2_1$ cubic phase to a low symmetry modulated structure, while the ferromagnetic transition takes place at 370K [4]. An interesting aspect of Ni-Mn-Ga alloys is the isothermal giant entropy change obtained when structural and magnetic transition temperatures nearly coincide, leading to a development of new materials exhibiting magnetocaloric effect [5]. The latest candidates in the field of FSMA has been alloys with composition $Ni_2Mn_{2-x}Z_x$ with Z = In, Sn, Sb [6].

$Ni_{2+x}Mn_{1-x}Ga$ ($0 \leq x \leq 0.19$) displays monotonic increase in martensitic transformation temperature, $T_M$ and a decrease in ferromagnetic ordering temperature, $T_C$ with increasing Ni concentration until both merge at x = 0.19 [7]. This has been attributed to increasing electron per atom (e/a) ratio. The crystal structure of all these alloys in the austenitic phase is ordered $L2_1$ while the low temperature crystal structure consists of different intermartensitic transformations as the lattice is subjected to periodic shuffling of the (110) planes along the [1-10]P direction of the initial cubic system [8] with modulation period dependent on the composition as summarized in [9]. Recent calculations by [10, 11] indicate the importance of modulated structure and the shuffling of atomic planes in stabilizing the martensitic structure.

$Ni_2MnSn$, that is isostructural to $Ni_2MnGa$ but with slightly higher electron per atom (e/a) ratio does not exhibit any structural instability leading to martensitic transition. However substitution of Mn for Sn induces a martensitic transition in the similar temperature range as that of preceding series. Of particular interest is $Ni_2Mn_{2-x}Sn_x$ with $0.5 < x \leq 0.6$ for which an inverse magnetocaloric effect is observed that is nearly three times larger in comparison to other alloys [12]. $Ni_2Mn_{1.4}Sn_{0.6}$ orders ferromagnetically at a Curie temperature $T_C$ = 319 K while martensitic transition occurs at temperature, $T_M$ = 200 K [6]. Neutron diffraction experiments on $Ni_2Mn_{1.36}Sn_{0.56}$ show that the cubic $L2_1$ structure in the austenitic phase transforms to orthorhombic 4O structure with Pmma space group in the martensitic phase [13]. Further, the magnetic moment of $Ni_2Mn_{1.36}Sn_{0.56}$ in the martensitic phase is smaller by about 50% than that in cubic phase [14]. Even in the $L2_1$ phase, the Mn moments are significantly smaller than those reported for the stoichiometric $Ni_2MnZ$ alloys [13,15]. It is conjectured that apart from the ferromagnetic order, some antiparallel alignment of the excess Mn moments could exist in $Ni_2Mn_{2-x}Sn_x$ [13,16]. Further, interesting aspect is the higher $T_M$ in $Ni_2Mn_{1.4}In_{0.6}$ alloy as compared to its Sn counterpart [17].

In spite of intense efforts, the underlying mechanism giving rise to such a phase transformation is still not well understood. The nature of modulations forming the super structures and the driving force for the martensitic transformation in these alloys is currently at debate. An understanding at the microscopic level of such transformation can be achieved by making a comparative study of the local structure in going from austenitic to martensitic phase. Thus a precise knowledge of the changes occurring in the tetragonal sub-unit of the martensitic phase is fundamental in understanding the mechanism involved in martensitic transformations.

**Experimental**

All the samples were prepared by repeated melting of the appropriate quantities of the constituent elements of 4N purity under argon atmosphere in an arc furnace. The sample beads so obtained were sealed in evacuated quartz ampoules and annealed at 800 K for 48 h followed by quenching in cold water. Energy dispersive x-ray analyses were performed to confirm the composition of the samples. The sample beads were cut and thoroughly ground to a very fine powder for x-ray diffraction (XRD) and EXAFS measurements while a small piece of the same bead was used for magnetization study. The room temperature crystal structure was determined by XRD recorded on Rigaku D-MAX IIC diffractometer with Cu Kα radiation. The magnetization measurements were carried out on a Vibrating Sample Magnetometer in the low field value (50 Oe) and an a.c. susceptometer in the temperature range 50 to 350 K. EXAFS at Ni and Mn and Ga K-edge were recorded at room temperature and liquid nitrogen temperature in the transmission mode on the EXAFS-1 beamline at ELETTRA Synchrotron Source using Si(111) as monochromator. Data analyses were carried out using IFEFFIT in ATHENA and ARTEMIS programs [18,19].

**Results**

**$Ni_{2+x}Mn_{1-x}Ga$**

Figure 1 shows the evolution of $T_M$ and $T_C$ as a function of excess Ni concentration (x). The two transition temperatures were determined from a.c. susceptibility measurements [20]. It can be seen that with the increase in Ni concentration at the expense of Mn, $T_M$ increases while $T_C$ decreases until they merge at 334K for x = 0.19.

Magnitude of $k^3$ weighted Fourier transform (FT) spectra for Mn and Ga K-edge EXAFS in austenitic $Ni_2MnGa$, martensitic $Ni_{2.16}Mn_{0.84}Ga$ and non-modulated $Ni_{2.19}Mn_{0.81}Ga$ are presented in figure 2. The first peak at around R = 2.5 Å is due to scattering from the nearest-neighbour shell comprising of 8 Ni atoms. Martensitic transition being diffusionless, no drastic variation at least in the first shell is expected in the two phases. Indeed, it is seen from figure 2 that the first peak position remains unchanged in the FT spectra of Ga K EXAFS for all the three samples. However, in case of FT of Mn K EXAFS a shift to higher R in the position of the first peak is observed for the sample in martensitic phase. This is very important observation in context to modulated structures in martensitic phase. Further, the broad peaks in the range R = 2.8 - 5.0 Å are due to the combined contribution from the second to fourth single scattering paths and some relatively weak multiple scattering paths. In this region of R, a difference in spectral signatures of the three compositions is quite evident. This can be attributed to the lowering of symmetry from the parent cubic structure due to martensitic transition.

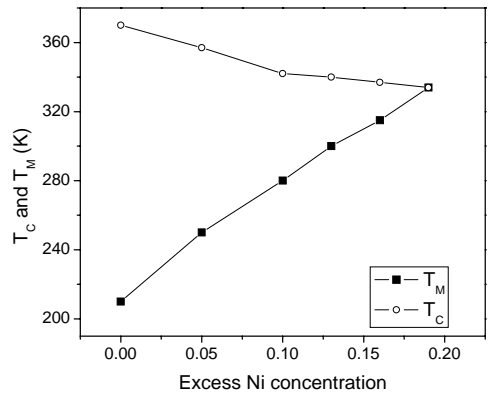

Figure 1 Variation of $T_M$ and $T_C$ as a function of excess Ni concentration (x) in $Ni_{2+x}Mn_{1-x}Ga$

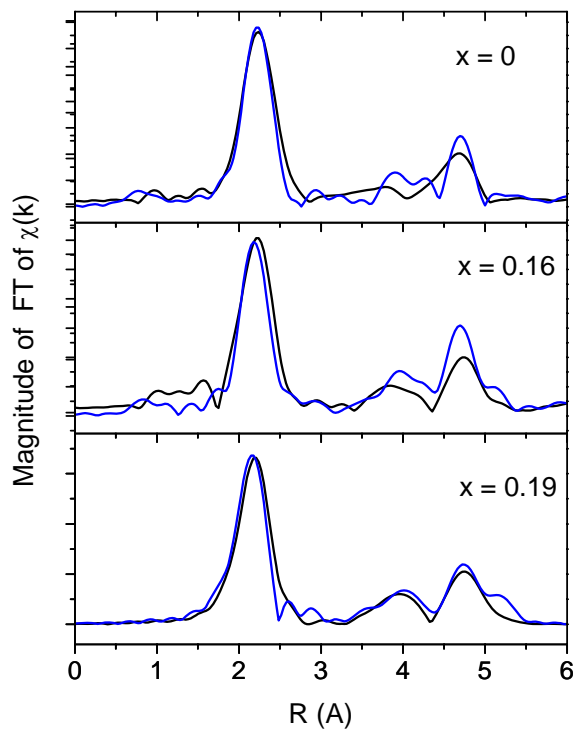

Figure 2 The $k^3$ weighted FT EXAFS spectra taken at 300K for Mn and Ga K-edge for $Ni_2MnGa$, $Ni_{2.16}Mn_{0.84}Ga$ and $Ni_{2.19}Mn_{0.81}Ga$. An indication of change in the nearest neighbour bond-length for $Ni_{2.16}Mn_{0.84}Ga$ is evident from the shift of the first peak (~2.5Å) in Ga K-edge spectra (differently coloured curve).

The Fourier transform of $k^3$ weighted EXAFS spectra was fitted in the R range of 1 to 5Å with the appropriate structural model using the first 4 single scattering and one linear multiple scattering correlation. The details of the model and fitting procedure have been explained in ref. [21]. The fitting for Mn and Ga EXAFS are presented in figure 3 and the best fit parameters are presented in Tables I and II.

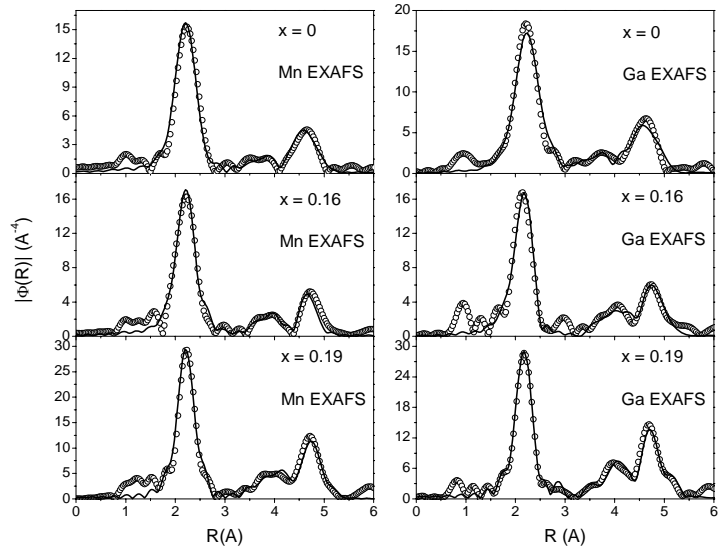

**Figure 3** Magnitude of Fourier transform of Mn and Ga K edge EXAFS in $Ni_{2+x}Mn_{1-x}Ga$ for x = 0, 0.16 and 0.19. The solid line is the best fit to the data.

**Table I** Results of the fits to Mn edge EXAFS in $Ni_{2+x}Mn_{1-x}Ga$ (x = 0, 0.16, 0.19). Figures in parentheses indicate uncertainty in the last digit.

| Atom and Co-ord. No. | x = 0 | | x = 0.16 | | x = 0.19 | |
|---|---|---|---|---|---|---|
| | R(Å) | $\sigma^2$(Å$^2$) | R(Å) | $\sigma^2$(Å$^2$) | R(Å) | $\sigma^2$(Å$^2$) |
| Ni X 8 | 2.519(8) | 0.0081(3) | 2.523(2) | 0.0051(2) | 2.521(2) | 0.0081(2) |
| Ga X 2 | 2.909(3) | 0.03(1) | 2.739(3) | 0.0054(3) | 2.739(5) | 0.0093(5) |
| Ga X 4 | | | 3.23(2) | 0.012(3) | 3.23(4) | 0.03(1) |
| Mn X 4 | 4.114(4) | 0.029(9) | 3.89(3) | 0.008(4) | 3.93(3) | 0.019(5) |
| Mn X 8 | | | 4.23(1) | 0.011(1) | 4.24(1) | 0.012(1) |
| Ni X 16 | 4.824(5) | 0.019(3) | 4.61(1) | 0.009(2) | 4.60(1) | 0.019(2) |
| Ni X 8 | | | 5.327(8) | 0.0044(8) | 5.360(8) | 0.0134(4) |
| MS X 16 | 5.038(5) | 0.0097(6) | 5.075(4) | 0.0068(4) | 5.082(5) | 0.0122(6) |

MS = multiple scattering path, for eg. Mn→Ni→Ga

**Table II** Results of the fits to Ga edge EXAFS in $Ni_{2+x}Mn_{1-x}Ga$ (x = 0, 0.16, 0.19). Figures in parentheses indicate uncertainty in the last digit.

| Atom and Co-ord. No. | X = 0 | | X = 0.16 | | X = 0.19 | |
|---|---|---|---|---|---|---|
| | R(Å) | $\sigma^2$(Å$^2$) | R(Å) | $\sigma^2$(Å$^2$) | R(Å) | $\sigma^2$(Å$^2$) |
| Ni X 8 | 2.512(2) | 0.0077(2) | 2.511(2) | 0.0042(1) | 2.512(2) | 0.0074(2) |
| Mn X 2 | 2.901(2) | 0.030(6) | 2.722(3) | 0.0067(5) | 2.710(5) | 0.0086(6) |
| Mn X 4 | | | 3.0(2) | 0.03(3) | 3.24(4) | 0.016(9) |
| Ga X 4 | 4.103(3) | 0.022(4) | 3.85(2) | 0.009(2) | 3.85(3) | 0.016(4) |
| Ga x 8 | | | 4.248(7) | 0.009(1) | 4.27(1) | 0.011(1) |
| Ni x 16 | 4.811(4) | 0.015(1) | 4.619(7) | 0.0083(6) | 4.62(3) | 0.023(4) |
| Ni x 8 | | | 5.319(8) | 0.0025(5) | 5.36(1) | 0.007(1) |
| MS x 16 | 5.025(3) | 0.014(1) | 5.106(2) | 0.0047(3) | 5.131(1) | 0.0104(6) |

MS = multiple scattering path, for eg. Ga →Ni→Mn

The important finding of the entire analysis is the change that occurs in the first shell of the central atom. The variation in the Mn-Ni and Ga-Ni bond distances obtained from the Mn and Ga edges in the three samples reveal a lot about the underlying martensitic transformations. From the

Mn K-edge data analysis it is seen that the Mn-Ni bond distance increases from 2.51Å to 2.52Å in going from the austenitic (x = 0) to martensitic (x = 0.16) structure. However, from the Ga K-edge data analysis the Ga-Ni bond length does not follow the same trend. As seen from the Table I and II, Ga-Ni distance remains constant at 2.512Å. Furthermore, if one considers the difference between Mn-Ni and Ga-Ni bond distances in martensitic phase (x = 0.16) alone, this change is amplified to 0.012Å. Both the central atoms, Mn and Ga can be viewed to be at the body centred position of a reduced tetragonal structure formed by Ni atoms. Non-uniformity in their bond distance with Ni of the order of $10^{-2}$Å is unexpected and hints towards the microscopic changes influencing the formation of the macroscopic modulated phases. Furthermore, for x = 0.16 sample a discrepancy between Mn-Ga and Ga-Mn distances is observed. Ideally, these distances should have been same as they involve the same pair of atoms. However as can be seen from Tables I and II, Mn-Ga distance (third bond distance) obtained with Mn as central atoms is 3.23Å while the same bond distance, with Ga as central atom comes out to be 3.0Å. Also the $\sigma^2$ values obtained from Ga EXAFS for Ga-Mn correlations are larger than those obtained from Mn EXAFS. This discrepancy in bond distances vanishes for x = 0.19 sample which is reported to have non modulated structure [22]. The physical significance of these observations is that the Ga atoms have smaller amplitude of displacement from its crystallographic position in comparison to Mn. In other words, Ga atoms are sluggish and do not get much displaced in undergoing a martensitic transition leading to a stronger hybridization between Ga and Ni in the martensitic phase. This directly points towards some sort of structural distortions in the martensitic phase. The disagreement in the value of bond lengths obtained from Ga K-edge analysis and from those obtained from Mn edge analysis based on the orthorhombic structural model can be understood if one considers the movement of Ga atom in the (110) plane along the [1-10]P direction. A modulated structure with Ga atom moved by 1% from its crystallographic position of (0, 0, 1/2), in the cubic phase can fully explain the observed bond-lengths.

It is this movement of Ga atoms that gives rise to modulations and the associated superstructures while the rigid Mn atoms forms an orthorhombic cell. It may be noted that the room temperature crystal structure of $Ni_{2.16}Mn_{0.84}Ga$ has been reported in literature to be of 7M modulated type [23]. It appears from EXAFS analysis that when the sample undergoes martensitic transition, Ga and Ni atoms move closer to each other while the Mn atoms move away. A similar conclusion has also been drawn from the first principles calculation of electronic structure by Zayak et al [11]. Such a movement of atoms results in increase in $p - d$ hybridization at the Fermi level. Zayak et al [10] have shown that the enhancement of interaction between Ga p and Ni d results in a dip in minority spin DOS at the Fermi level which is responsible for martensitic phenomena in Ni-Mn-Ga alloys.

**$Ni_2Mn_{1.4}Z_{0.6}$ (Z = Sn and In)**

In figure 4 X-ray diffraction patterns of $Ni_2Mn_{1.4}Z_{0.6}$ (Z = Sn and In) are shown. It can be seen that for the Sn sample, the diffraction pattern clearly indicates an ordered $L2_1$ structure. The lattice parameter obtained from fitting this pattern is 5.9941±0.0003Å. In the case of ordered $Ni_2Mn_{1.4}Sn_{0.6}$, Ni occupies the X site at (¼,¼,¼), the Y site at (0,0,0) is occupied by Mn and the Z site at (½,½,½) is occupied by Sn(60%) and Mn(40%). However in the case of In sample, the diffraction pattern is slightly different. This could be due to the martensitic transformation temperature of 295K for this sample. In fact in literature this composition is reported to crystallize in different structures viz, 10M and orthorhombic respectively [17,24]. Although the diffraction pattern corresponding to In sample in figure 4 shows all cubic reflections, it also shows a shoulder on the higher angle side of the main peak (see inset in figure 4) which is indicative of lowering of structural symmetry due to martensitic transformation. With reference to the diffraction patterns presented in ref. 24, the diffraction pattern is indexed in Pm3m space group giving a lattice constant of 3.019±0.001Å.

EXAFS data at the Mn and Ni K edges at RT and 100K were fitted with the respective structural models as obtained from XRD analysis.

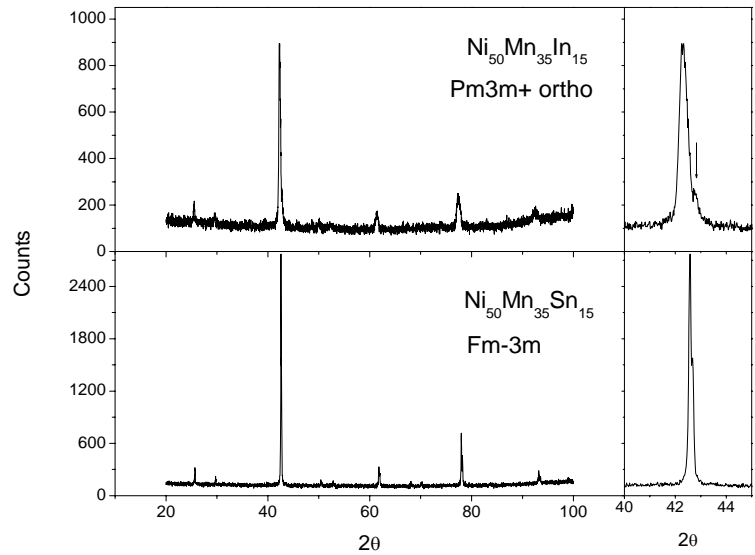

Figure 4 X-ray diffraction plots of $Ni_2Mn_{1.4}Sn_{0.6}$ and $Ni_2Mn_{1.4}In_{0.6}$. The insets show the cubic (220) reflection.

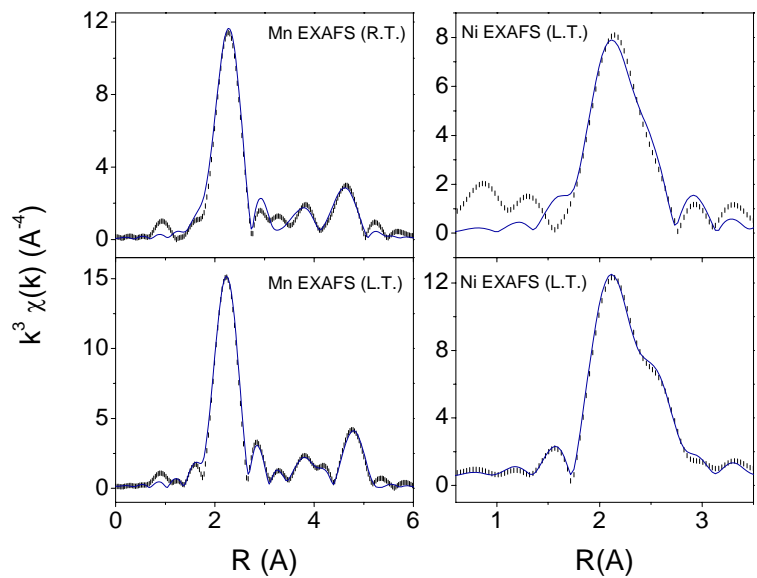

Figure 5 Magnitude of Fourier transform of Mn and Ni K edge EXAFS at RT and 100K in $Ni_2Mn_{1.4}Sn_{0.6}$. The solid line is the best fit to the data.

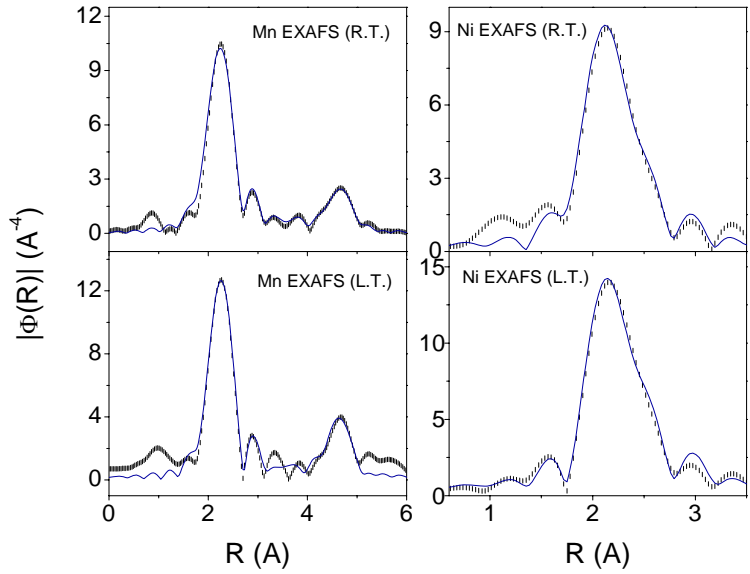

**Figure 6 Magnitude of Fourier transform of Mn and Ni K edge EXAFS at RT and 100K in $Ni_2Mn_{1.4}In_{0.6}$. The solid line is the best fit to the data.**

The fittings in R space in the range 1 to 5Å for Mn EXAFS and 1 to 3Å for Ni EXAFS are presented in figures 5 and 6 for $Ni_2Mn_{1.4}Sn_{0.6}$ and $Ni_2Mn_{1.4}In_{0.6}$ respectively. The best fit parameters at RT (austenitic) and 100K (martensitic) are presented in Table III. The details of the analysis are reported in ref. [25] and [26].

The EXAFS measurements in the cubic (austenitic) and martensitic phase of $Ni_2Mn_{1.4}Sn_{0.6}$ bring out an important observation that influences the martensitic transition in this system. Essentially, the present system belongs to the Ni-Mn based ternary Heusler intermetallics having generic formula $X_2YZ$. The stoichiometric $Ni_2MnSn$ is a ferromagnet with $T_C \sim 360K$ but does not undergo any martensitic transformation [15]. The exchange interactions in such Heusler alloys are mainly due to the indirect RKKY interaction between Mn atoms mediated by the conduction electrons of the system [27]. As Mn concentration is increased in the system, martensitic phase transformation sets in and magnetism gets even more complex. The magnetization results reported in the previous section show that $Ni_2Mn_{1.4}Sn_{0.6}$ orders ferromagnetically at $T_C = 319$ K and undergoes a martensitic transformation $\sim 200$ K.

The important result of the EXAFS study is that the local distortions exist in the crystal structure within the cubic framework. These distortions reflect through a shorter Ni-Mn distance and distinctly different Ni-Sn and Ni-Mn bond lengths at RT. The Ni-Mn distance obtained from EXAFS analysis is the average bond length of Ni-[Y-Mn] and Ni-[Z-Mn]. Since the atomic sizes of Sn and Mn are different, a local distortion can occur when Mn atoms occupy the Sn sites. This can lead to a shorter Ni-[Z-Mn] distance as compared to Ni-[Y-Mn] distance. This distortion in the $L2_1$ structure may be one of the factors that influence the martensitic transformation in $Ni_2Mn_{1.4}Sn_{0.6}$. A shorter Ni-[Z-Mn] bond implies stronger hybridization of Ni with Z-Mn. The hybridization features between X and Z species of the $X_2YZ$ metallic systems is known to influence the binding mechanism [28]. In the case of $Ni_2Mn_{1.4}Sn_{0.6}$, the stronger Ni-[Z-Mn] hybridization perhaps results in a redistribution of electrons causing changes in the DOS at Fermi level leading to a martensitic transition. In $Ni_2Mn_{1.4}In_{0.6}$ the crystal structure being B2, there is a complete disorder between the Y and Z sites and the identification of Mn that hybridizes with Ni, as Mn at Y-site or Mn at Z-site is not possible. However, the fact that there exists some Mn-d and Ni-d hybridization is clear from the shorter Ni-Mn and longer Ni-In bonds at RT as well as 100K. It is this $d-d$ hybridization that is responsible for martensitic transformation in $Ni_2Mn_{1.4}In_{0.6}$.

**Table III Results of the fits to Ni and Mn edge EXAFS in $Ni_2Mn_{1.4}Z_{0.6}$ (Z = Sn and In). Figures in parentheses indicate uncertainty in the last digit.**

| Atom and Coord. No. | R (Å) | $\sigma^2$(Å$^2$) | Atom and Coord. No | R (Å) | $\sigma^2$(Å$^2$) |
|---|---|---|---|---|---|
| $Ni_2Mn_{1.4}Sn_{0.6}$ | | | | | |
| Mn EXAFS at RT | | | Ni EXAFS at RT | | |
| Ni x 8 | 2.549(6) | 0.0127(7) | Mn x 5.6 | 2.550(7) | 0.0132(9) |
| Mn x 3.6 | 2.93(2) | 0.012(2) | Sn x 2.4 | 2.601(7) | 0.0072(6) |
| Sn x 2.4 | 2.95(1) | 0.008(1) | Ni x 6 | 2.98(3) | 0.026(5) |
| Mn XAFS at 100K | | | Ni EXAFS at 100K | | |
| Ni x 8 | 2.568(1) | 0.0077(1) | Mn x 5.6 | 2.569(2) | 0.0092(2) |
| Mn x 3.6 | 2.873(3) | 0.0069(3) | Sn x 2.4 | 2.607(1) | 0.0035(1) |
| Sn x 2.4 | 2.877(4) | 0.0095(5) | Ni x 2 | 2.83(1) | 0.016(2) |
| | | | Ni x 4 | 3.15(3) | 0.027(5) |
| $Ni_2Mn_{1.4}In_{0.6}$ | | | | | |
| Mn EXAFS at RT | | | Ni EXAFS at RT | | |
| Atom and Coord. No. | R (Å) | $\sigma^2$(Å$^2$) | Atom and Coord. No | R (Å) | $\sigma^2$(Å$^2$) |
| Ni x 8 | 2.567(3) | 0.0115(3) | Mn x 5.6 | 2.580(4) | 0.0132(9) |
| Mn x 4.2 | 2.886(5) | 0.0058(5) | In x 2.4 | 2.634(4) | 0.0072(6) |
| In x 1.8 | 2.911(5) | 0.0108(6) | Ni x 6 | 3.13(6) | 0.026(5) |
| Mn EXAFS at 100K | | | Ni EXAFS at 100K | | |
| Ni x 8 | 2.568(3) | 0.0115(3) | Mn x 5.6 | 2.558(3) | 0.0068(4) |
| Mn x 3.6 | 2.879(6) | 0.0058(5) | Sn x 2.4 | 2.70(1) | 0.009(1) |
| Sn x 2.4 | 2.896(6) | 0.0108(6) | Ni x 2 | 2.797(7) | 0.038(8) |
| | | | Ni x 4 | 3.13(8) | 0.027(13) |

A martensitic transformation occurs when the Fermi surface touches the Brillouin zone boundary [29]. This implies that change in factors like chemical pressure (as a result of difference in atomic sizes) and the e/a value can cause the alteration of the Fermi surface driving such systems towards structural instabilities. A linear dependence of the $T_M$ on changing e/a has indeed been observed for the $Ni_{2+x}Mn_{1-x}Ga$ alloys upto x = 0.19 [22]. However, in the present case, $Ni_2Mn_{1.4}Sn_{0.6}$ that has an e/a value of 8.05 undergoes a martensitic transition at much lower temperature as compared to $Ni_2Mn_{1.4}In_{0.6}$ with an e/a = 7.9. One of the main differences between these two systems is the crystal structure in the austenitic phase. $Ni_2Mn_{1.4}Sn_{0.6}$ has a cubic $L2_1$ structure while the $Ni_{50}Mn_{35}In_{15}$ crystallizes in a B2 structure. This could be perhaps due to the size difference between In and Mn in $Ni_2Mn_{1.4}In_{0.6}$ and Sn and Mn atoms in $Ni_2Mn_{1.4}Sn_{0.6}$. The size difference between In and Mn atoms being larger, results in a greater amount of disorder and therefore a more disordered B2 structure. Therefore, at least in case of Ni-Mn-In and Ni-Mn-Sn systems the martensitic transition temperatures depend upon the structural disorder rather than e/a ratio. This disorder also affects the hybridization in the martensitic phase. In this Sn containing sample the Ni-Mn and Ni-Sn bond lengths obtained from EXAFS analysis are 2.57°A and 2.61°A respectively, giving a difference of 0.04°A. Whereas as it can be seen from Table III the difference between Ni-Mn and Ni-In bond lengths is 0.14°A. This clearly indicates that the local structural disorder in $Ni_2Mn_{1.4}In_{0.6}$ results in a stronger Ni(3d)-Mn(3d) hybridization in the martensitic phase. Such hybridization is responsible for a rearrangement of the d electrons within the hybrid band resulting in lifting of degeneracy and lowering of the symmetry. This being stronger in the case of $Ni_2Mn_{1.4}In_{0.6}$ as compared to $Ni_2Mn_{1.4}Sn_{0.6}$ results in higher martensitic transition temperature inspite of lower e/a ratio.

**Conclusions**

- In $Ni_2MnGa$, increased hybridization and distortion of tetrahedral symmetry leads to re-distribution of electrons causing band Jahn-Teller effect.

- In $Ni_2Mn_{1.4}Z_{0.6}$ disorder in the austenitic (cubic) phase is responsible for the system being unstable towards structural transformation.

- There is a change in the Ni-Mn hybridization upon martensitic phase transformation.

- This hybridization is stronger in $Ni_2Mn_{1.4}In_{0.6}$ sample due to larger Mn tetrahedral distortions resulting in a higher martensitic transformation temperature.

**Acknowledgements**

Financial assistance from Council of Scientific and Industrial Research (CSIR) New Delhi is gratefully acknowledged. Thanks are due to Department of Science and Technology, Government of India and ICTP-Elettra for travel assistance and local hospitality for XAFS experiments.